\documentclass[conference]{IEEEtran}
\IEEEoverridecommandlockouts
\usepackage{cite}
\usepackage{amsmath,amssymb,amsfonts}
\usepackage{algorithmic}
\usepackage{graphicx}
\usepackage{textcomp}
\usepackage{xcolor}
\def\BibTeX{{\rm B\kern-.05em{\sc i\kern-.025em b}\kern-.08em
    T\kern-.1667em\lower.7ex\hbox{E}\kern-.125emX}}

    \usepackage{scalerel}
\usepackage{tikz}
\usetikzlibrary{svg.path}

\definecolor{orcidlogocol}{HTML}{A6CE39}
\tikzset{
  orcidlogo/.pic={
    \fill[orcidlogocol] svg{M256,128c0,70.7-57.3,128-128,128C57.3,256,0,198.7,0,128C0,57.3,57.3,0,128,0C198.7,0,256,57.3,256,128z};
    \fill[white] svg{M86.3,186.2H70.9V79.1h15.4v48.4V186.2z}
                 svg{M108.9,79.1h41.6c39.6,0,57,28.3,57,53.6c0,27.5-21.5,53.6-56.8,53.6h-41.8V79.1z M124.3,172.4h24.5c34.9,0,42.9-26.5,42.9-39.7c0-21.5-13.7-39.7-43.7-39.7h-23.7V172.4z}
                 svg{M88.7,56.8c0,5.5-4.5,10.1-10.1,10.1c-5.6,0-10.1-4.6-10.1-10.1c0-5.6,4.5-10.1,10.1-10.1C84.2,46.7,88.7,51.3,88.7,56.8z};
  }
}

\newcommand\orcidlink[1]{\href{https://orcid.org/#1}{\mbox{\scalerel*{
\begin{tikzpicture}[yscale=-1,transform shape]
\pic{orcidlogo};
\end{tikzpicture}
}{|}}}}

\usepackage[hidelinks]{hyperref}
    
\begin{document}

\title{Analysis of Library Dependency Networks of Package Managers Used in iOS Development
\thanks{This research has been funded by grant PRG1226 of the Estonian Research Council, the European Social Fund via the IT Academy program, and
the Federal Ministry for Climate Action, Environment, Energy, Mobility, Innovation and Technology (BMK), the Federal Ministry for Digital and Economic Affairs (BMDW), and the State of Upper Austria in the frame of the SCCH competence center INTEGRATE (FFG grant no. 892418) part of the COMET - Competence Centers for Excellent Technologies Programme managed by Austrian Research Promotion Agency FFG.
%the Austrian ministries BMVIT and BMDW and the Province of Upper Austria in the frame of the SCCH COMET competence center INTEGRATE.
}
}

\author{\IEEEauthorblockN{1\textsuperscript{st} Kristiina Rahkema \orcidlink{0000-0001-7332-2041}
}
\IEEEauthorblockA{
\textit{Institute of Computer Science} 
\\
\textit{University of Tartu}
\\
Tartu, Estonia 
\\
kristiina.rahkema@ut.ee}
\and
\IEEEauthorblockN{2\textsuperscript{nd} Dietmar Pfahl \orcidlink{0000-0003-2400-501X}
}
\IEEEauthorblockA{\textit{Institute of Computer Science} 
\\
\textit{University of Tartu}
\\
Tartu, Estonia 
\\
dietmar.pfahl@ut.ee}
\and
\IEEEauthorblockN{3\textsuperscript{rd} Rudolf Ramler \orcidlink{0000-0001-9903-6107}
}
\IEEEauthorblockA{\textit{Software Competence Center Hagenberg}
\\
\textit{(SCCH) GmbH}
\\
Hagenberg, Austria
\\
rudolf.ramler@scch.at}}

\maketitle

\begin{abstract}
Reusing existing solutions in the form of third-party libraries is common practice when writing software. 
Package managers are used to manage dependencies to third-party libraries by automating the process of installing and updating the libraries. Library dependencies themselves can have dependencies to other libraries creating a dependency network with several levels of indirections.
The library dependency network in the Swift ecosystem encompasses libraries from CocoaPods, Carthage and Swift Package Manager (PM). These package managers are used when developing, for example, iOS or Mac OS applications in Swift and Objective-C.  We provide the first analysis of the library dependency network evolution in the Swift ecosystem.

Although CocoaPods is the package manager with the biggest set of libraries, the difference to other package managers is not as big as expected. The youngest package manager and official package manager for Swift, Swift PM, is becoming more and more popular, resulting in a gradual slow-down of the growth of the other two package managers. 
When analyzing direct and transitive dependencies, we found that the mean total number of dependencies is lower in the Swift ecosystem compared to many other ecosystems. Still, the total number of dependencies shows a clear growing trend over the last five years.  
\end{abstract}


\section{Introduction}
Reusing existing solutions in the form of third-party libraries is common practice when writing software. This makes the development process faster and easier. And, third-party solutions are often better vetted than custom solutions.

Using a package manager allows to declare and keep track of a project's dependencies to third-party libraries. The library dependencies themselves can, again, have dependencies to other libraries, creating a network of library dependencies. The collection of all libraries that are available through a package manager and their library dependencies create a - potentially large and complex - library dependency network (LDN). The structure and evolution of such LDNs of various package managers have been studied. For example, Kikas et al.~\cite{kikas2017structure} created a dependency dataset and analyzed the LDNs of JavaScript, Ruby and Rust. Decan et al.~\cite{decan2019empirical} studied the growth of LDNs of seven package managers. They found that the number of libraries, versions and dependencies grew for each package manager linearly or even exponentially, increasing the risk of dependency conflicts and incompatibilities. 

Furthermore, when the number of direct and transitive dependencies (i.e. indirect dependencies of any level of indirection) grows, it also increases the security risk of a library depending on vulnerable library version. As seen in the example of Log4J, even a vulnerability in a seemingly harmless logging library can have a severe impact on a significant part of a software ecosystem~\cite{log4j}. 

Although several studies analyzed LDNs, especially for npm and Maven, no studies exist for the LDNs of CocoaPods, Carthage and Swift PM. These three package managers are used when developing applications in Swift, such as iOS, Mac OS or Watch OS applications. In the following, we refer to the combined ecosystems of CocoaPods, Carthage and Swift PM as the Swift ecosystem. It is important to note that this ecosystem contains libraries written in other languages as well (e.g., Objective-C, C, C++). Additionally, CocoaPods and Carthage are also used in many Objective-C projects. A further interesting aspect of the Swift ecosystem is that the LDNs of the three package managers are partially overlapping. 

In this paper, we present our analysis of the evolution of the LDNs of CocoaPods, Carthage and Swift PM. We analyse the growth of the whole Swift ecosystem in terms of number of libraries and number of library versions, the growth of LDNs for each of the three package managers, and the evolution of the number of dependencies over time. By comparing our results and observations to studies of LDNs of other ecosystems, we also share insights related to the characteristics of iOS development. We previously briefly discussed these research plans in a doctoral symposium paper \cite{rahkema2022quality}.

\section{Background}

\subsection{Package Managers}
Our focus is on libraries that can be used in applications written in Swift, such as iOS or Mac OS applications. Package managers used in Swift development are CocoaPods, Carthage, and Swift Package Manager (Swift PM).

\textbf{CocoaPods}\footnote{\url{https://cocoapods.org}} was released in September 2011 and is the oldest package manager with around 88,000 libraries. It is a centralized package manager. Dependencies are declared in a Podfile. When CocoaPods is executed, it downloads and compiles the declared libraries. It generates a new Xcode Workspace with all libraries included. This makes CocoaPods easy to use, as there is no additional manual work needed. 

\textbf{Carthage}\footnote{\url{https://github.com/Carthage/Carthage}} was released in November 2014. According to Libraries.io, it includes 4.5 thousand libraries~\cite{librariesio}. This number, however, is an estimate as Carthage is a decentralized package manager and no official central repository of libraries exists. Carthage was created as a counterweight to the more heavyweight CocoaPods. Libraries can be included through Carthage by simply adding a repository address of a library to the Cartfile. Carthage downloads and compiles these libraries but does not automatically include them in the app projects. 

\textbf{Swift Package Manager (Swift PM)}\footnote{\url{https://www.swift.org/package-manager/}} was released in December 2017. It is the official package manager created by Apple. Swift PM is a decentralized package manager like Carthage. Differently to the other two package managers, the Package.swift file is also used as a build file. Support for iOS applications was not added to Swift PM until 2019~\cite{swiftpm_ios}. Since 2019 it is also possible to use Swift PM directly through Xcode, the main IDE for iOS and Mac OS development. 

\subsection{Dependency Data}

For each package manager there exist configuration files that specify which libraries the developers wish to include in their project. There are two types of configuration files: manifest files and resolution files. Developers specify a library and its version constraints in a manifest file. After installing the dependencies through the package manager, a resolution file is generated by the package manager that specifies the exact library versions installed. Each of the package managers has a slightly different way of declaring dependencies, the underlining idea however is the same. 

\subsection{LDN Dataset}

Rahkema et al.~\cite{rahkema-msr} created a dataset for the LDNs of the Swift ecosystem. The dataset contains information on library versions, their dependencies, and publicly reported vulnerabilities. The dataset contains data on 60,533 libraries, 572,131 library versions, and 23,419 dependencies between libraries. We use this data on library versions and dependencies to analyse the evolution of the LDNs.  

\section{Related Work}

In related work, the evolution of LDNs for various package managers has been analyzed, most often for Maven, npm and RubyGems. So far, no studies exist on package managers in the Swift ecosystem (CocoaPods, Carthage and Swift PM). 

Kikas et al.~\cite{kikas2017structure} analyzed the evolution of LDNs of three languages: JavaScript, Ruby and Rust. They found that for each package manager the number of libraries is growing. Similarly, the number of direct dependencies and total dependencies per project is increasing. The increase was extreme for JavaScript, where the average number of total dependencies grew from one per project to almost 60 between 2011 and 2016.

Decan et al. analyzed the evolution of seven package manager LDNs~\cite{decan2019empirical}. They used the libraries.io dataset to analyze how these package managers' LDNs change over time. They found that the growth of the number of libraries and dependencies differs between package managers. Some LDNs grow linearly while others grow exponentially. They showed that the number of transitive dependencies is significantly higher than the number of direct dependencies. For some of the package managers, the ratio between transitive and direct dependencies is growing. They also pointed out that the average dependency depth is between three and six, depending on the package manager. The libraries.io data set includes data about CocoaPods, Carthage and Swift Package Manager (the three package managers used in iOS development), but according to Decan et al. this data was incomplete, i.e. there was data on dependencies between library versions for these package managers. Therefore, these package managers were excluded from the analysis. 

Kula et al. analyzed dependency updates in 4,600 java projects~\cite{kula2018developers}. They found, that 81.5\% of the studied projects did not update their outdated dependencies. They plotted library usage curves and discovered that new library versions are mostly used by new dependent projects only. 

\section{Method}

Scripts used in our analyses can be found on GitHub\footnote{\url{https://github.com/kristiinara/LibraryDependencyEvolution}}.
In this paper, we analyse the following research questions: 

\begin{itemize}
    \item RQ1: How has the combined LDN of the Swift ecosystem evolved? 
    \item RQ2: How have the LDNs of each of the package manager evolved? 
    \item RQ3: How has the number of dependencies evolved in the LDNs? 
\end{itemize}

\subsection{RQ1: Evolution of the combined LDN}

For RQ1, we plot the cumulative number of all libraries and library versions including libraries that have no dependencies and no dependents. This might include unused libraries. 

\subsection{RQ2: Evolution of the LDNs}

For RQ2, we only consider connected libraries, i.e., libraries with at least one dependent or dependency. 
We first look at how the number of libraries has grown for each package manager. For this, we find the first version of each library, group by the month of its commit timestamp, and count the number of unique libraries cumulatively. We plot the cumulative curve for each package manager. 
We then calculate how the number of library versions has grown for each package manager. Again, we group the library versions by the months of commit timestamps. We count the number of library versions released each month and take the cumulative sum. The cumulative curve is plotted for each package manager. 

\subsection{RQ3: Evolution of Dependencies}

For RQ3, we plot the mean number of direct and transitive dependencies for each month, as a monthly snapshot. The monthly snapshot is calculated by finding library versions released during each month and for each library taking the last library version for each month. 

The number of direct and transitive dependencies is found by querying LIBRARY\_DEPENDS\_ON chains with a length of up to 10. A maximum threshold for the dependency chain needs to be set for performance reasons, we did, however, confirm that very few dependencies existed beyond that level. The ratio of the total number of dependencies to the number of libraries is then calculated and plotted. 

\section{Results and Discussion}

\subsection{RQ1: Evolution of the combined LDN}

\begin{figure}
  \includegraphics[width=1.0\columnwidth]{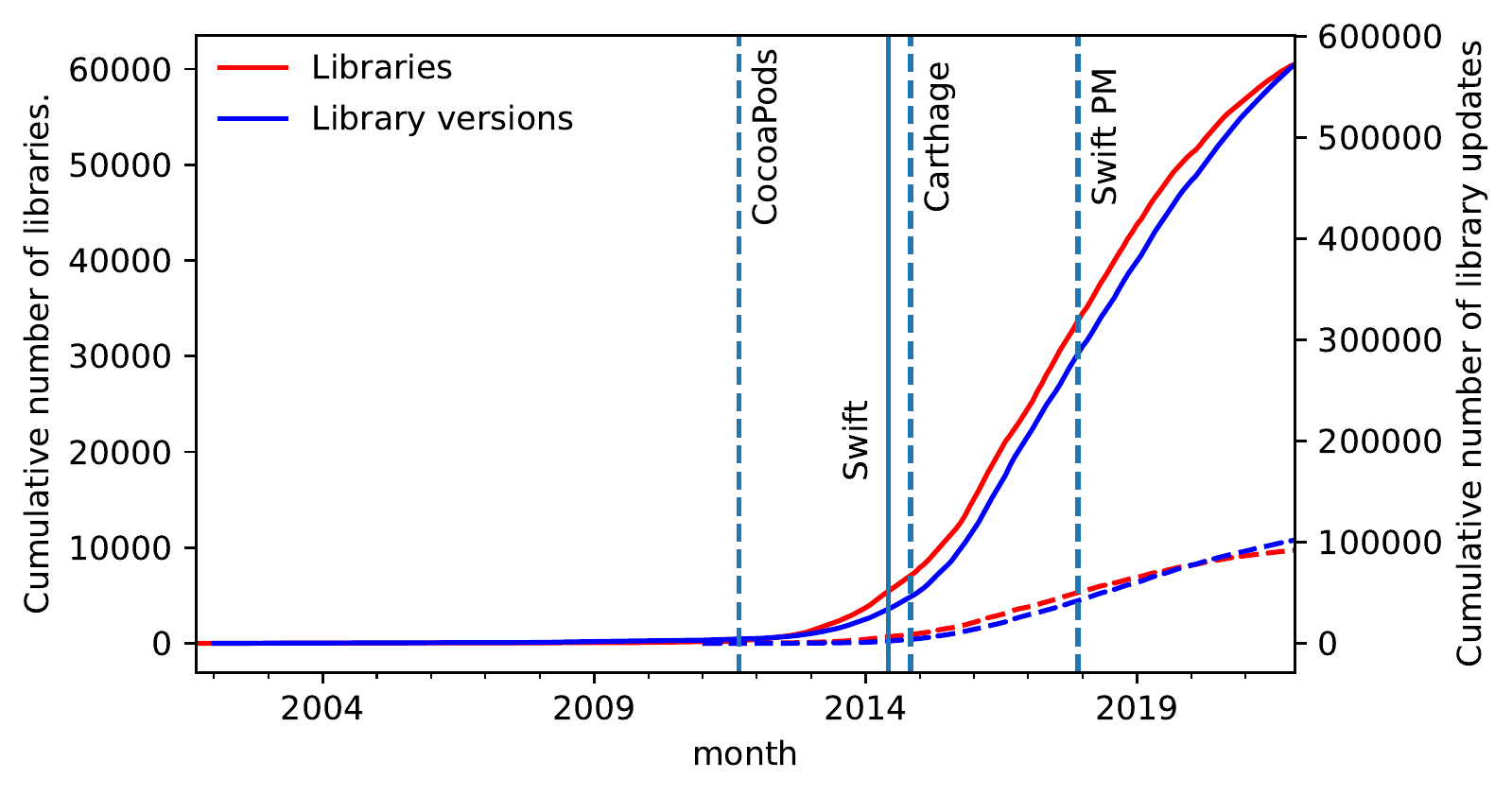}
  \caption{Cumulative number of libraries and library versions. Solid lines show numbers for all libraries, dotted lines for connected libraries.}
  \label{fig:libraries-cumulative}
\end{figure}

We analyzed 60,533 libraries in total. Figure~\ref{fig:libraries-cumulative} shows the cumulative number of libraries (red) and library versions (blue) over time. The subset of connected libraries and their versions is shown as dotted lines. The total number of libraries grew very fast after the release of the Swift programming language in 2014. From 2019 onward the number of new libraries added has slightly slowed down. A similar pattern can be observed for library versions. Moreover, we see similar trends for the subset of connected libraries, i.e., libraries that either use a package manager or are used through a package manager.

From Figure \ref{fig:libraries-cumulative} it is evident that some of the libraries in the Swift ecosystem were created before the introduction of the first package manager, CocoaPods, or before the release of the language Swift. Firstly, it is important to note that, although we call it the Swift ecosystem, it also encompasses Objective-C libraries. Objective-C, the predecessor of Swift, was introduced in 1984 and it is inter-operable with Swift. Additionally, libraries written in C and C++ can be used in both Swift and Objective-C projects. Some of these libraries are also available through these package managers. 

Some of the C/C++/Objective-C libraries written before the release of CocoaPods were later added to the package managers. In our analysis we use git tags and commit timestamps to date library versions. Therefore, as we have no information on when a library or library version was added to a package manager, we simply assume that it was added when the library version was released. In the following, we will ignore any library versions added before September 2011, i.e., prior to when CocoaPods was introduced.

\subsection{RQ2: Evolution of the LDNs}

In the following analysis, we constrain ourselves to libraries that have at least one dependency or dependent. This means that these libraries are indeed part of at least one package manager LDN. These libraries are called connected libraries. 

We analyzed the cumulative number of connected libraries for each package manager. In total there are 9,755 connected libraries. Of these libraries, 6,600 belonged to the CocoaPods LDN, 2,856 belonged to Carthage and 2,150 belonged to Swift PM. A library can belong to multiple package manager LDNs.

\begin{figure}
  \includegraphics[width=1.0\columnwidth]{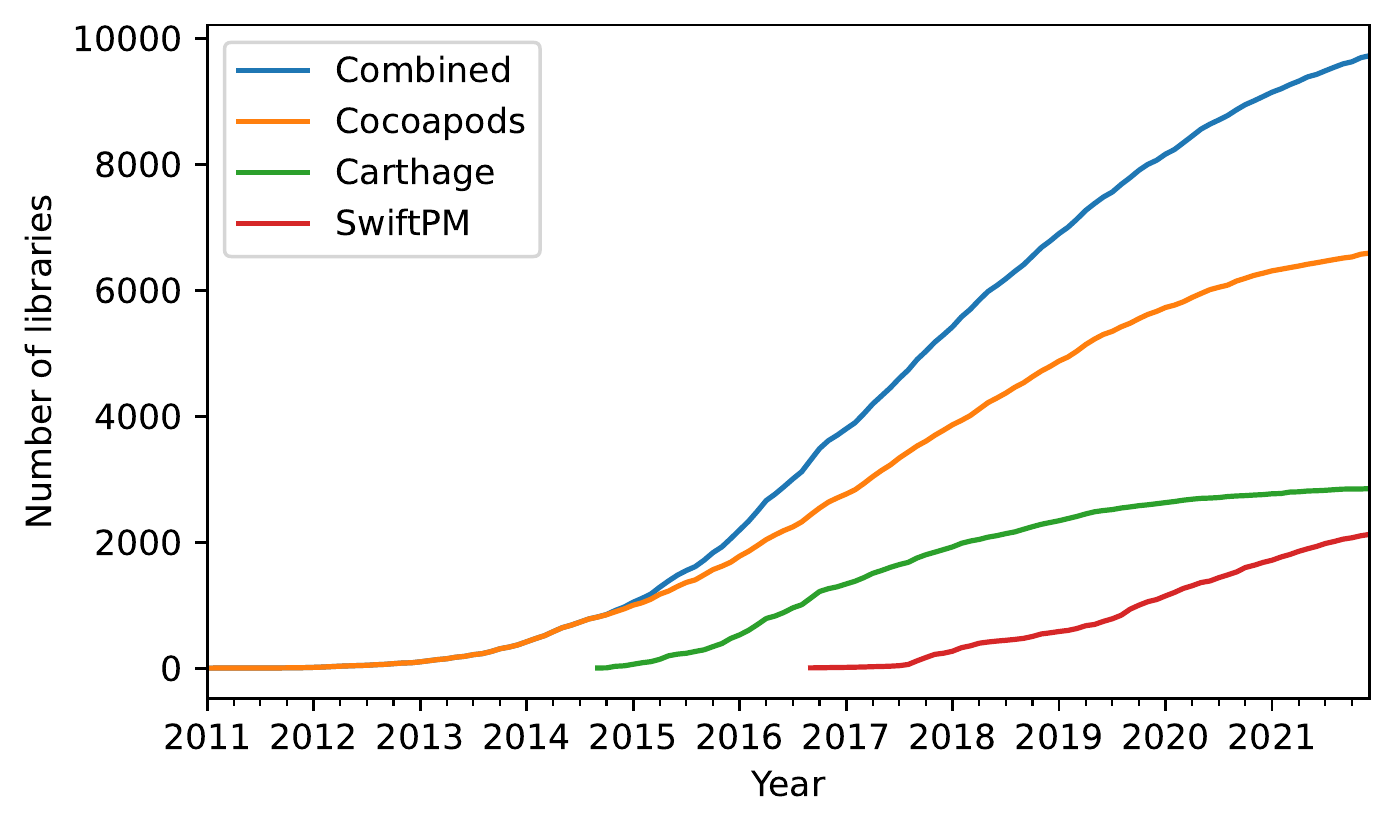}
  \caption{Cumulative number of libraries.}
  \label{fig:libraries-pm-cumulative}
\end{figure}

The change in the number of libraries can be seen in Figure~\ref{fig:libraries-pm-cumulative}. The number of libraries is growing fastest for the newest and smallest package manager Swift PM. The number of libraries for CocoaPods is still growing, but the growth has slowed after 2019. The growth of the number of libraries for Carthage has almost completely halted.

When Decan et al.~\cite{decan2019empirical} analyzed the LDNs of seven package managers, they found that the number of libraries grew for each package manager either linearly or exponentially. Given these results we expected to see a similar trend in the number of libraries and library versions for CocoaPods, Carthage and Swift PM.  
However, the growth we observed has slowed in recent years. The decline in growth for CocoaPods and Carthage is also clear when looking at the number of new library versions added per month.

To check if the number of updates per month for CocoaPods is really declining, we conducted an additional analysis of the git history of CocoaPods Spec repository. In this analysis, we searched for the latest git commit for each month and then queried the difference between commits of consecutive months. Additions and deletions were recorded for each file. We discarded all file names that were not Podspec files and then counted the number of additions for each month. This analysis confirmed that the number of updates (i.e., file additions) were indeed falling. 

Our hypothesis is that more and more developers are moving to Swift PM. Therefore, Carthage has lost most of its appeal. First, when Apple introduced Swift PM it was a standalone terminal application that could be used to create Mac OS applications and packages. Furthermore, in 2019, Apple added support for iOS and Xcode, which is the official IDE for iOS and Mac OS development. Now a dependency can be declared through Swift PM by simply searching for a library in Xcode. This makes Swift PM the easiest to use package manager in the Swift ecosystem. 

\subsection{RQ3: Evolution of Library Dependencies}

\begin{figure}
  \includegraphics[width=1.0\columnwidth]{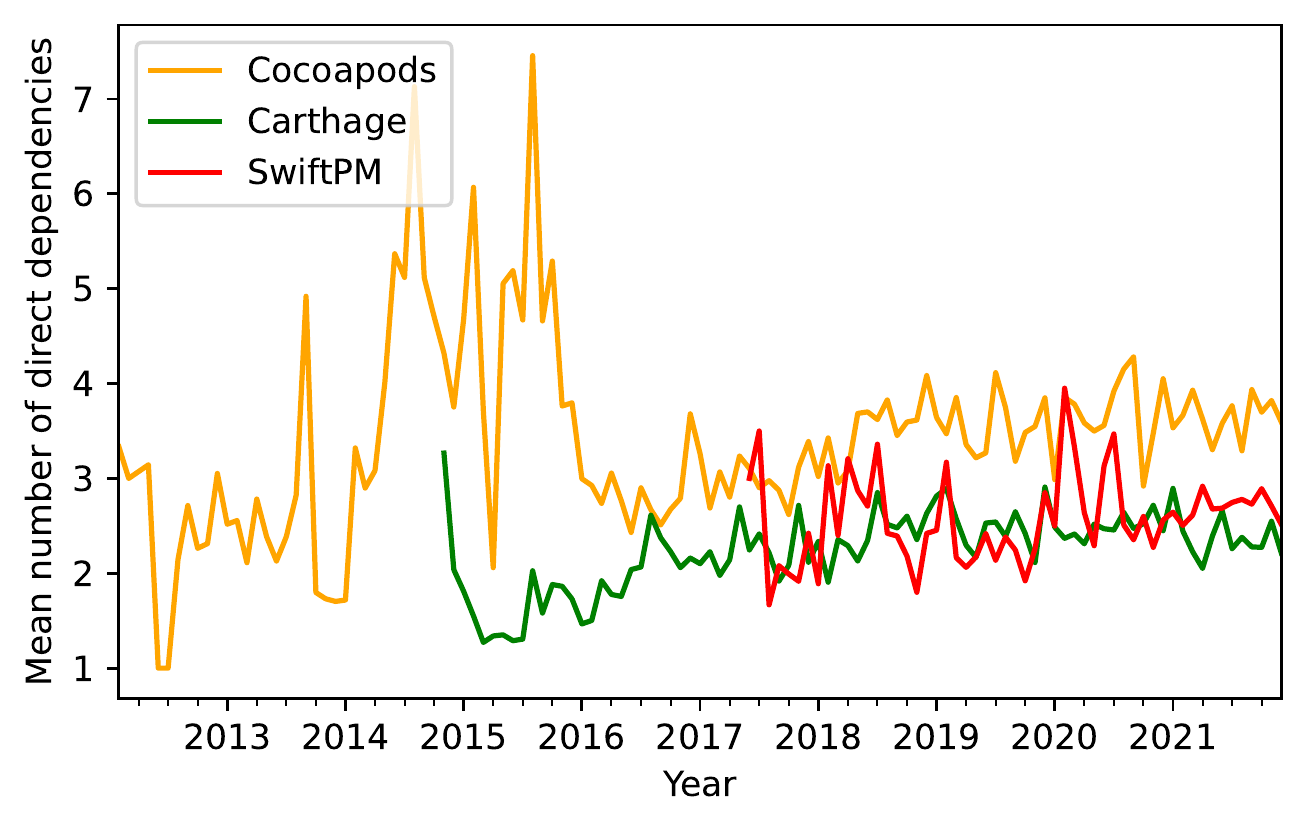}
  \caption{Number of direct dependencies for each monthly snapshot.}
  \label{fig:dependencies}
\end{figure}

%Next, we looked at the evolution of library dependencies. 
Figure \ref{fig:dependencies} shows the mean number of direct dependencies for each monthly snapshot. For CocoaPods, the mean number of direct dependencies fluctuated strongly until 2016. After 2016 the mean number leveled to values around three, which is slightly higher than the mean number of direct dependencies for Carthage and Swift PM, each averaging around 2.5. The mean number of direct dependencies has a slight upwards trend for all three package managers. 

\begin{figure}
  \includegraphics[width=1.0\columnwidth]{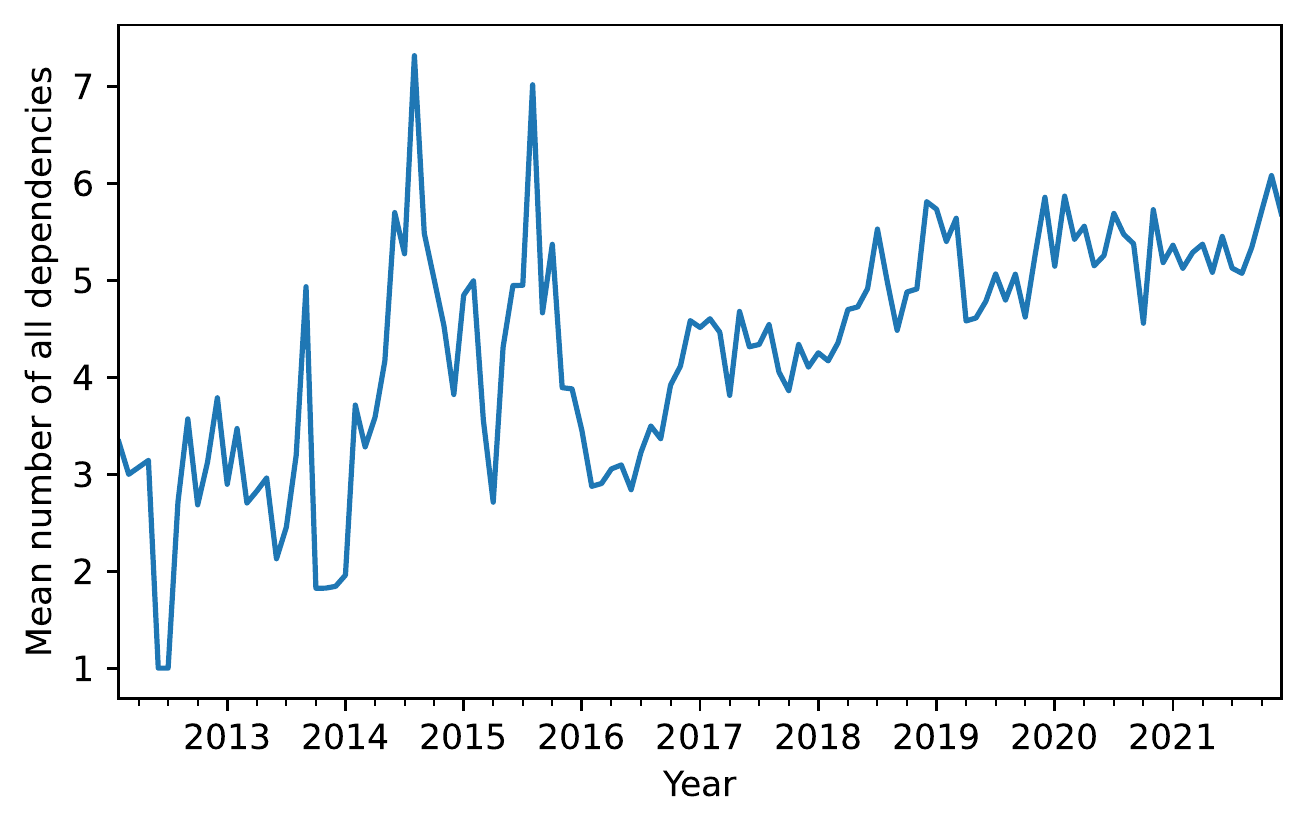}
  \caption{Total number of dependencies for each monthly snapshot.}
  \label{fig:dependencies-all}
\end{figure}

In addition, we calculated the mean number of direct and transitive dependencies for all connected libraries. The data shown in Figure~\ref{fig:dependencies-all} is not differentiating between package managers, as calculating transitive dependency chains conditional to package managers was difficult. We did, however, count the number of unique library names as total dependencies in order to not accidentally include the same library twice if it was referenced through multiple package managers. The mean number of dependencies in Figure~\ref{fig:dependencies-all} shows a clear upwards trend. Similarly to the mean number of direct dependencies, the number of all dependencies fluctuates considerably until 2016. Between 2016 and 2022, however, there is a clear upwards trend where the mean number of direct and transitive dependencies rises from around 3 to 5.5.  

We expected the number of direct dependencies and total dependencies to grow over time as has been observed by Kikas et al.~\cite{kikas2017structure}. We observed that the number of direct dependencies was very volatile and growing rapidly between 2012 and 2016. After 2016 the number of direct dependencies stabilized and started growing slowly. Although the mean number of direct dependencies is consistently higher for CocoaPods than for Carthage, the trend is similar for all three package managers. Our hypothesis is that this change was a consequence of the introduction of Swift in 2014 and developers migrating from Objective-C to Swift in the following years.

The growth of the mean number of dependencies after 2016 for all three studied package managers is lower than for JavaScript, Ruby and Rust~\cite{kikas2017structure}. The mean number of direct dependencies is comparable to other package managers~\cite{decan2019empirical}, but the mean number of transitive dependencies is significantly lower. For example, the median number of transitive dependencies for a library available through the Cargo, NuGet and npm package manager is 41, 27 and 21, respectively. In comparison, the median number of transitive dependencies in the LDNs of the Swift ecosystem is only two. 

\section{Conclusion}

We analysed the combined LDN of the three package managers used in the Swift ecosystem: CocoaPods, Carthage and Swift PM. We saw that the cumulative number of libraries and library versions is growing, but has slowed down in recent years. Swift PM, the newest and only official package manager for Swift, seems to be on the rise as it is growing faster than the other package managers during the last couple of years. 

In comparison to LDNs of the Cargo, NuGet or npm ecosystems, the Swift ecosystem shows a smaller number of library dependencies. The resulting benefit may be a smaller probability of depending on a vulnerable library.

Overall, the number of direct dependencies is showing a slight upwards trend, while the number of total dependencies is clearly growing, reminding of Lehman's law of ever increasing complexity \cite{1456074}.

\bibliographystyle{IEEEtran}
\bibliography{bibliography}

\end{document}